\documentclass[12pt,thmsa]{article}
\usepackage{amsmath}
\usepackage{graphicx}
\usepackage{amsfonts}
\usepackage{amssymb}

\begin{document}

\begin{center}
{\Huge Cartan's topological structure}
\end{center}

\vspace{1pt}

\begin{center}
\qquad\textbf{R. M. Kiehn and Phil Baldwin}

Physics Department, University of Houston

http://www.cartan.pair.com\vspace{1pt}\bigskip
\end{center}

\begin{quote}
\textbf{\vspace{1pt}Abstract:}A system of differential forms will establish a
topology and a topological structure on a domain of independent variables such
that is possible to determine which maps or processes acting on the system are
continuous. \ Perhaps the most simple topology is that generated by the
existence of a single 1-form of Action, its Pfaff sequence of exterior
differentials, and their intersections. \ \ In such a topology the exterior
derivative becomes a limit point generator in the sense of Kuratowski. \ \ The
utilization of such techniques in physical systems is examined. \ A key
feature of the Cartan topology is determined by the Pfaff dimension
(representing the minimum number of functions to describe the 1-form
generator). \ In particular, when the Pfaff dimension is 3 or more the Cartan
topology becomes a disconnected topology, with the existence of topological
torsion and topological parity. \ Most classical physical applications are
constrained to cases where the Pfaff dimension is 2 or less, for such is the
domain of unique integrability. \ The more interesting domain of non-unique
solutions requires the existence of topological torsion, and can lead to an
understanding of irreversible processes without the use of statistics.
\end{quote}

\section{Introduction}

\qquad In the period from 1899 to 1926, Eli Cartan developed his theory of
exterior differential systems \lbrack1,2\rbrack, which included the ideas of
spinor systems \lbrack3\rbrack\ and the differential geometry of projective
spaces and spaces with torsion \lbrack4\rbrack. The theory was appreciated by
only a few contemporary researchers, and made little impact on the main stream
of physics until about the 1960's. Even specialists in differential geometry
(with a few notable exceptions \lbrack5\rbrack\ ) made little use of Cartan's
methods until the 1950's. Even today, many physical scientists and engineers
feel that Cartan's theory of exterior differential forms is just another
formalism of fancy.

However, Cartan's theory of exterior differential systems has several
advantages over the methods of tensor analysis that were developed during the
same period of time. The principle fact is that differential forms are well
behaved with respect to functional substitution of C1 differentiable maps.
Such maps need not be invertible even locally, yet differential forms are
always deterministic in a retrodictive sense [6], by means of functional
substitution. Such determinism is not to be associated with contravariant
tensor fields, if the map is not a diffeomorphism.

Although the word ''topology'' had not become popular when Cartan developed
his ideas (topological ideas were described as part of the theory of analysis
situs), there is no doubt that Cartan's intuition was directed towards a
topological development. For example, Cartan did not define what were the open
sets of his topology, nor did he use in his early works the words ''limit
points or accumulation points'' explicitly, but he did describe the union of a
differential form and its exterior derivative as the ''closure$^{"}$of the form.

In a simplistic comparison it might be said that tensor methods are restricted
to geometric applications, while Cartan's methods can be applied directly to
topological concepts as well as geometrical concepts. Cartan's theory of
exterior differential systems is a topological theory not necessarily limited
by geometrical constraints and the class of diffeomorphic transformations that
serve as the foundations of tensor calculus. A major objective of this article
is to show how limit points, intersections, closed sets, continuity,
connectedness and other elementary concepts of modern topology are inherent in
Cartan's theory of exterior differential systems. These ideas do not depend
upon the geometrical ideas of size and shape (hence Cartan's theory, as are
all topological theories, is renormalizeable). In fact the most useful of
Cartan's ideas do not depend explicitly upon the geometric ideas of a metric,
nor upon the choice of a connection as in fiber bundle theories.

In this article the Cartan topology will be constructed explicitly for an
arbitrary exterior differential system, $\Sigma.$ \ All elements of the
topology will be evaluated, and the limit points, the boundary sets and the
closure of every subset will be computed abstractly. An earlier intuitive
result [7]\ utilized the notion that Cartan's concept of the exterior product
may be used as an intersection operator, and his concept of the exterior
derivative may used as a limit point operator acting on differential forms.
These ideas will be given formal substance in this article. A major result of
this article, with important physical consequences in describing evolutionary
processes, is the demonstration that the Cartan topology is not necessarily a
connected topology, unless the property of topological torsion vanishes.

\subsection{The exterior product}

Cartan's theory of exterior differential systems has its foundations in the
Grassmann algebra, where the two combinatorial processes used to define the
algebra are: vector space addition, and what is now called the exterior
product [8]. The exterior product acts on pairs of algebraic elements called
exterior p-forms. In this article, the exterior or wedge product operator is
symbolized by the \symbol{94}\ symbol for ease of typing; the exterior product
of $A$ and $B$ is then given by the expression $A\symbol{94}B$. The p-forms
may have a differential basis, symbolized by the set $dx$, combined in the
manner of a vector space with functional coefficients. A differential 1-form
is then given by the expression,%

\begin{equation}
A=A_{\mu}dx^{\mu}.
\end{equation}
For 1-forms the multiplication rules are:%

\begin{equation}
A\symbol{94}A=0,\,A\symbol{94}B=-B\symbol{94}A.
\end{equation}

At some regular point, $\{x\}$, the 1-form will admit N-1 vector fields,
$\mathbf{V}$, to be constructed such that%

\begin{equation}
i(\mathbf{V})A=A_{\mu}V^{\mu}=0.
\end{equation}
Such vector fields are defined to be ''associated'' vectors of the 1-form, $A
$. \ If these vector fields have a vanishing Lie bracket, then they span a
neighborhood of the point in terms of a simple (hyper) surface. The adjoint
vector to this N-1 system acts as the ''normal'' field to the surface spanned
by the N-1 vectors. The coefficients $A_{\mu}$~ form this ''normal'' field.

Now consider two such surface systems represented by the 1-forms $A$ and $B$.
Do the two (curved) surfaces intersect? The points in common to the two
surfaces are given by the non-null set formed by the exterior product of $A$
with $B$. If $A\symbol{94}B$ vanishes, then the two surfaces have no points in
common. Consider the simple case where the 1-forms $A$ and $B$ have
coefficients which form the components of a gradient field ( a Gauss
Weingarten surface normal),%

\begin{equation}
A=A_{\mu}dx^{\mu}=\nabla\phi\cdot d\mathbf{r\,\,\,\,\,\,\,\,\,\,\,\,\,\,\,}%
B=B_{\mu}dx^{\mu}=\nabla\psi\cdot d\mathbf{r}%
\end{equation}

Then ( in 3 dimensions ) $A\symbol{94}B=(\mathbf{A}\times\mathbf{B}%
)_{z}dx\symbol{94}dy\,\,$+ cyclic permutations), a result that demonstrates
that the Gibbs cross product, $\mathbf{A}\times\mathbf{B}=\nabla\phi
\times\nabla\psi$ in euclidean three dimensions is related to the ''line'' of
points which are in common to both surfaces. \ This result pictorially cements
the notion that the exterior product (acting on 1-forms) is an operator
related to the concept of intersection. If the two surfaces do not intersect,
and the exterior product vanishes, then the functions $\phi$ and $\chi$ are
not functionally independent. \ These concept extends to p-forms of higher rank.

\ In three dimensions, the Gibbs cross product is considered to be a
''vector'' for it has the same number of components as the gradient. \ Yet it
has different behavior under transformations of the basis, and is sometimes
called a ''pseudovector'' or an axial vector. \ In the exterior calculus, the
exterior product of the two 1-forms, with components proportional to covariant
tensor of rank 1, creates a 2-form with covariant components of rank 2. \ Only
in constrained geometries, such as euclidean three space, do 2-forms have any
resemblence to the Gibbs cross product (a rule which fails in dimension n
$>$%
3). \ The pseudo-vector is an object that behaves like a contravariant tensor
density of rank 1. \ \ Such objects are usually defined as ''currents''.
\ \ In general, there are two species of differential forms (that are often
dual to one another and are well behaved with respect to functional
substitution and the pullback operation: p-forms and N-p form densities or
currents. One species pulls back (meaning that the form is well defined with
respect to functional substitution) with respect to the Jacobian transpose,
while the other pulls back with respect to the Jacobian adjoint. \ Of course
for orthogonal systems, these concepts are degenerate, for the inverse and the
adjoint and the transpose of the Jacobian matrix are the same. \ Recall that
at a point it is always possible to define a vector basis in terms of an
orthogonal system (use the Gram-Schmidt process), but it may not be possible
to extend the orthogonality concept smoothly (without singularities) from
\ one neighborhood to another neighborhood. \ If the neighborhoods can be
connected by a singly parameterized vector field, then these concepts are at
the basis of the Frenet-Serret moving frame analysis. \ Cartan extended these
ideas to domains that are not so simply connected, and developed the notion of
the moving basis Frame, which he called the Repere Mobile.

\subsection{The exterior derivative}

\vspace{1pt}The second new operator found in Cartan's theory of exterior
differential systems is the exterior derivative. The exterior derivative, like
the exterior product, also has topological connotations when applied to
differential forms, but the results are sometimes surprising and unfamiliar.
For example consider the exterior derivative of the N-1 form density, $D$, in
three dimensions, given by the expression,%

\begin{align}
dD  &  =d(D^{x}dy\symbol{94}dz-D^{y}dz\symbol{94}dx+D^{z}dx\symbol{94}dy)\\
&  =div_{3}(\mathbf{D})dx\symbol{94}dy\symbol{94}dz\Rightarrow\rho
(x,y,z)dx\symbol{94}dy\symbol{94}dz\nonumber
\end{align}
where $\rho$ has been defined as the resultant of the action of the exterior
derivative, $div_{3}(\mathbf{D})$. The usual interpretation of Gauss' law is
that the field lines of the vector $\mathbf{D}$ terminate (or have a limit or
accumulation point) on the charges, $Q$. \ \ Gauss' law generates both the
intuitive idea that sources are related to limit points, and the novel concept
that the exterior derivative is a limit point operator creating these limits
points when the operation is applied to a differential form. However, as
demonstrated below, the concept that the exterior derivative is a limit point
operator relative to the Cartan topology is a general idea, and is not
restricted to Gauss' law.

For example, extending this idea to four dimensions for the N-2 form density,
$G$, of Maxwell excitations ($\mathbf{D,H}$), \ %

\begin{equation}
\vspace{1pt}G=-D^{x}dy\symbol{94}dz+D^{y}dz\symbol{94}dx-D^{z}dx\symbol{94}%
dy+H^{x}dx\symbol{94}dt+H^{y}dy\symbol{94}dt+H^{z}dz\symbol{94}dt,
\end{equation}
the exterior derivative $dG$ of $G$ yields a three form, $J$, defined as the
electromagnetic current 3-form,%

\begin{equation}
\vspace{1pt}J=J^{x}dy\symbol{94}dz\symbol{94}dt-J^{y}dx\symbol{94}%
dz\symbol{94}dt+J^{z}dx\symbol{94}dy\symbol{94}dt-\rho dx\symbol{94}%
dy\symbol{94}dt
\end{equation}
where in 3-vector language,%

\begin{equation}
curl\,\,\mathbf{H}-\partial\mathbf{D}/\partial
t=0\,\,\,\,\,\,\,\,\,\,\,\,\,\,\,\,\,div\,\mathbf{D}=\rho.
\end{equation}
The charge current density act as the ~''limit points'' of the Maxwell field
excitations. \ Note that $dJ=0$ for C2 functions by Poincare's lemma.

However, consider the N-1 current, $C$ (not necessarily equal to $J$ as
defined above) in four dimensions%

\begin{equation}
C=\rho\{V^{x}dy\symbol{94}dz\symbol{94}dt-V^{y}dx\symbol{94}dz\symbol{94}%
dt+V^{z}dx\symbol{94}dy\symbol{94}dt-1dx\symbol{94}dy\symbol{94}dt\}
\end{equation}
and its exterior derivative as given by the expression,%

\begin{equation}
dC=\{div_{3}(\rho\mathbf{V})+\partial\rho/\partial t\}dx\symbol{94}%
dy\symbol{94}dz\symbol{94}dt.=Rdx\symbol{94}dy\symbol{94}dz\symbol{94}%
dt=R\,\Omega_{4\_vol}%
\end{equation}

When thre 4-form $R$ vanishes, the resultant expression is physically
interpreted as the ''equation of continuity'' or as a ''conservation law''.
Over a closed boundary, that which goes in is equal to that which goes out
(when $dC=0$). \ Note that the concept of the conservation law is a
topological constraint: the ''limit points'' of the ''current 3-form'' in four
dimensions must vanish if the conservation law is to be true. If the RHS of
the above expression is not zero, then the current 3-form is said to have an
''anomaly'', or a source (or sink) . The anomaly acts as the source of the
otherwise conserved quantity. \ The limit points, $R$, of the 3-form, $C$, are
generated by its exterior derivative, $dC=\{div_{3}(\rho\mathbf{V}%
)+\partial\rho/\partial t\}\Omega_{4}.$ \ When the RHS is zero, the current
''lines'' do not stop or start within the domain. \ (It is possible for them
to be closed on themselves in certain topologies).

As another example, consider the 1-form of vector and scalar potentials given
by the expression,%

\begin{equation}
A=A_{x}dx+A_{y}dy+A_{z}dz-\phi dt=\mathbf{A~\cdot\,}d\mathbf{r}-\phi dt.
\end{equation}
The exterior derivative of the 1-form\thinspace\ $A$ generates the
2-form\thinspace\thinspace$F=dA$ of electromagnetic intensities,
($\mathbf{E,B}$):%

\begin{equation}
F=dA=B_{z}dx\symbol{94}dy+B_{x}dy\symbol{94}dz+B_{y}dz\symbol{94}%
dx+E_{x}dx\symbol{94}dt+E_{y}dy\symbol{94}dt+E_{z}dz\symbol{94}dt,
\end{equation}
where%

\begin{equation}
\mathbf{B}=curl\mathbf{A},\,\,\,\,\,\,\,and\,\,\,\,\,\mathbf{E}=-\nabla
\phi-\partial\mathbf{A}/\partial t.
\end{equation}
\vspace{1pt}The exterior derivative of $F$ vanishes if the potential functions
are C2 differentiable:%

\begin{equation}
ddA=dF=0\supset curl\mathbf{E}+\partial\mathbf{B}/\partial
t=0\,\,\,\,\,and\,\,\,\,\,div_{3}\mathbf{B}=0.
\end{equation}
Note that these derivations of Maxwell's equations are based on topological
statements about limit points, and do not depend upon geometrical
considerations of metric or connections.

This now almost classic generation of the Maxwell field equations [9]\ has
another less familiar interpretation: The $\mathbf{E}$ and$\,\,\mathbf{B}$
field intensities are the topological limit ''points'' of the 1-form of
potentials, $\{\mathbf{A},\phi\}$, relative to the Cartan topology! \ The
limit points of the 2-form of field intensities, $F$, are the null set. For C2
\ vector fields, the Cartan topology admits flux quanta, charge quanta, and
spin quanta, but excludes magnetic monopoles [10]. When the differential
system of interest is built upon the forms $A$, $F$ and $G$, it is possible to
show that superconductivity is to be associated with the constraints on the
limit point sets of $A,\,\,\,A\symbol{94}F,\,\,\,and\,\,\,A\symbol{94}%
G\,\,$[11]. That is, superconductivity has its origins in topological, not
geometrical, concepts. This remarkable idea that the exterior derivative is a
limit point operator is given formal substance in the section 4.

\section{The Cartan Point Set Topology.}

Cartan built his theory around an exterior differential system, ~$\Sigma$,
which consists of a collection of\thinspace\ 0- forms, 1-forms, 2-forms, etc.
[12]. He defined the closure of this collection as the union of the original
collection with those forms which are obtained by forming the exterior
derivatives of every p-form in the initial collection. In general, the
collection of exterior derivatives will be denoted by $d\Sigma$, and the
closure of $\Sigma$ by the symbol, $\Sigma^{c}$, where%

\begin{equation}
\Sigma^{c}=\Sigma\cup d\Sigma
\end{equation}

For notational simplicity in this article the systems of p-forms will be
assumed to consist of the single 1-form, $A$. Then the exterior derivative of
$A$ is the 2-form $F=dA$, and the closure of $A$ is the union of $A$ and
$F:A^{c}=A\cup F$. The other logical operation is the concept of intersection,
so that from the exterior derivative it is possible to construct the set
$A\symbol{94}F$\thinspace\thinspace\thinspace\thinspace defined collectively
as $H:H=A\symbol{94}F.$ The exterior derivative of $H$ produces the set
defined as $K=dH$, and the closure of \ $H$ is the union of $H$ and
$K:H^{c}=H\cup K$.

\qquad This ladder process of constructing exterior derivatives, and exterior
products, may be continued until a null set is produced, or the largest p-form
so constructed is equal to the dimension of the space under consideration. The
set so generated is defined as a Pfaff sequence. The largest rank of the
sequence determines the Pfaff dimension of the domain (or class of the form,
\lbrack13\rbrack), which is a topological invariant. The idea is that the
1-form $A$ (in general the exterior differential system, $\Sigma$) generates
on space-time four equivalence classes of points that act as domains of
support for the elements of the Pfaff sequence, $A,F,H,K$ . The union of all
such points will be denoted by $X=A\cup F\cup H\cup K$. The fundamental
equivalence classes are given specific names:%

\begin{equation}
Topo\log ical\,\,\,ACTION:A=A_{\mu}dx^{\mu}%
\end{equation}%

\begin{equation}
Topo\log ical\,\,\,\,VORTICITY:F=dA=F_{\mu\nu}dx^{\mu}\,\symbol{94}dx^{\nu}%
\end{equation}%

\begin{equation}
Topo\log ical\,\,\,TORSION:H=A\symbol{94}dA=H_{\mu\nu\sigma}dx^{\mu
}\,\symbol{94}dx^{\nu}\symbol{94}dx^{\sigma}%
\end{equation}%

\begin{equation}
Topo\log ical\,\,\,PARITY:K=dA\symbol{94}dA=K_{\mu\nu\sigma\tau}dx^{\mu
}\,\symbol{94}dx^{\nu}\symbol{94}dx^{\sigma}\symbol{94}dx^{\tau}.
\end{equation}

The Cartan topology is constructed from a basis of open sets, which are
defined as follows: first consider the domain of support of $A$. Define this
''point'' by the symbol $A$. $\ A$ is the first open set of the Cartan
topology. Next construct the exterior derivative, $F=dA$, and determine its
domain of support. Next, form the closure of $A$ by constructing the union of
these two domains of support, $A^{c}=A\cup F$. $\,\,A\cup F$ forms the second
open set of the Cartan topology. .

Next construct the intersection $H=A\symbol{94}F$, and determine its domain
$of$ support. Define this ''point'' by the symbol$\,\,H$, which forms the
third open set of the Cartan topology. Now follow the procedure established in
the preceding paragraph. Construct the closure of $H$ as the union of the
domains of support of $H$ and $K=dH$. The construction forms the fourth open
set of the Cartan topology. In four dimensions, the process stops, but for N
$>$%
4, the process may be continued.

Now consider the basis collection of open sets that consists of the subsets,%

\begin{equation}
B=\{A,\,A^{c},\,H,\,H^{c}\}=\{A,\,A\cup F,\,H,\,H\cup K\}
\end{equation}
The collection of all possible unions of these base elements, and the null
set, $\emptyset,$ generate the Cartan topology of open sets:%

\begin{equation}
\vspace{1pt}T(open)=\{X,\,\emptyset,\,A,\,H,A\cup F,\,H\cup K,\,A\cup
H,\,A\cup H\cup K,\,A\cup F\cup H\}.
\end{equation}
These nine subsets form the open sets of the Cartan topology constructed from
the domains of support of the Pfaff sequence constructed from a single 1-form,
$A$. \ The compliments of the open sets are the closed sets of the Cartan topology.%

\begin{equation}
\vspace{1pt}T(closed)=\{\emptyset,\,X,\,F\cup H\cup K,\,A\cup F\cup K,A\cup
F,\,\,H\cup K,F\cup K,F,\,K\}.
\end{equation}

From the set of 4 ''points'' $\{A,F,H,K\}$ that make up the Pfaff sequence it
is possible to construct 16 subset collections by the process of union. It is
possible to compute the limit points for every subset relative to the Cartan
topology. The classical definition of a limit point is that a point p is a
limit point of the subset Y relative to the topology T if and only if for
every open set which contains p there exists another point of Y other than p
[14]. The results of this definition are presented in Table I

\begin{center}
$%
\begin{array}
[c]{c}%
\text{\textbf{The Cartan Topology}}\\
A=A_{k}dx^{k}\\
F=dA,\;\ \ \ H=A\symbol{94}F,\;\ \ \ K=F\symbol{94}F\\
\text{Basis \{A, A}^{c}\text{, H, H}^{c}\text{\}}=\{A,A\cup F,H,H\cup K\}\\
T(open)=\{X,\,\emptyset,\,A,\,H,A\cup F,\,H\cup K,\,A\cup H,\,A\cup H\cup
K,\,A\cup F\cup H\}\\
T(closed)=\{\,\emptyset,X,F\cup H\cup K,A\cup F\cup K,\,H\cup K,\,A\cup
F,\,F\cup K,\,F,K\}
\end{array}
$
\end{center}

\vspace{1pt}

\begin{center}
\vspace{1pt}$%
\begin{array}
[c]{ccccc}%
\begin{array}
[c]{c}%
\text{\textbf{Subset}}\\
\sigma
\end{array}
&
\begin{array}
[c]{c}%
\text{\textbf{Limit Pts}}\\
d\sigma
\end{array}
&
\begin{array}
[c]{c}%
\begin{array}
[c]{c}%
\text{\textbf{Interior}}%
\end{array}
\\
.
\end{array}
&
\begin{array}
[c]{c}%
\text{\textbf{Boundary}}\\
\partial\sigma
\end{array}
&
\begin{array}
[c]{c}%
\text{\textbf{Closure}}\\
\sigma\cup d\sigma
\end{array}
\\
\emptyset & \emptyset & \emptyset & \emptyset & \emptyset\\
A & F & A & F & A\cup F\\
F & \emptyset & \emptyset &  F & F\\
H & K & H & K & H\cup K\\
K & \emptyset & \emptyset &  K & K\\%
\begin{array}
[c]{c}%
A\cup F\\
A\cup H\\
A\cup K\\
F\cup H\\
F\cup K\\
H\cup K
\end{array}
&
\begin{array}
[c]{c}%
F\\
F,K\\
F\\
K\\
\emptyset\\
K
\end{array}
&
\begin{array}
[c]{c}%
A\cup F\\
A\cup H\\
A\\
H\\
\emptyset\\
H\cup K
\end{array}
&
\begin{array}
[c]{c}%
\emptyset\\
F\cup K\\
F\cup K\\
F\cup K\\
F\cup K\\
\emptyset
\end{array}
&
\begin{array}
[c]{c}%
A\cup F\\
X\\
A\cup F\cup K\\
F\cup H\cup K\\
F\cup K\\
H\cup K
\end{array}
\\
A\cup F\cup H & F,K & A\cup F\cup K & K & X\\
F\cup H\cup K & K & H\cup K & F & F\cup H\cup K\\
A\cup H\cup K & F,K & A\cup H\cup K & F & X\\
A\cup F\cup K & F & A\cup F & K & A\cup F\cup K\\
X & F,K & X & \emptyset &  X
\end{array}
$
\end{center}

\vspace{1pt}By examining the set of limit points so constructed for every
subset of the Cartan system, and presuming that the functions that make up the
forms are C2 differentiable (such that the Poincare lemma is true,
$dd\omega=0,any\,\,\omega$), it is easy to show that for all subsets of the
Cartan topology every limit set is composed of the exterior derivative of the
subset, thereby proving the claim that the exterior derivative is a limit
point operator relative to the Cartan topology. \ \ For example, the open
subset, $A\cup H$, has the limit points that consist of \ $F$ and $K.\,$ The
limit set consists of $F\cup K$ which can be derived directly by taking the
exterior derivatives of the elements that make up $A\cup H$; that
is,$\,\,(F\cup A=d(A\cup H)=(dA\cup dH)$. Note that this open set, $A\cup H$,
does not contain its limit points. Similarly for the closed set, $A\cup F$,
the limit points are given by $F$ which may be deduced by direct application
of the exterior derivative to $(A\cup F):(F)=d(A\cup F)=(dA\cup dF)=(F\cup
\emptyset)=(F)$.\bigskip

\section{Topological Torsion and Connected vs Non-connected Cartan topologies.}

\qquad The Cartan topology as given in Table 1 is composed of the union of two
sub-sets which are both open and closed $(X=A^{c}\cup H^{c})$, a result that
implies that the Cartan topology is not necessarily connected. An exception
exists if the topological torsion, $H$, and hence its closure, vanishes, for
then the Cartan topology is connected. This extraordinary result has broad
physical consequences. The connected Cartan topology based on a vanishing
topological torsion is at the basis of most physical theories of equilibrium.
In mathematics, the connected Cartan topology corresponds to the Frobenius
integrability condition for Pfaffian forms. In thermodynamics, the connected
Cartan topology is associated with the Caratheodory concept of inaccessible
thermodynamic states \lbrack15\rbrack, and the existence of an equilibrium
thermodynamic surface. If the 1-form, $Q$, of heat generates a Cartan topology
of null topological torsion, $H=Q\symbol{94}dQ=\emptyset,$ then the Cartan
topology built on $Q$ is connected. Such systems are ''isolated'' in a
topological sense, and the heat 1-form has a representation in terms of two
and only two functions, conventionally written as: $Q=TdS$. Note again that a
fundamental physical concept, in this case the idea of equilibrium, is a
topological concept independent from geometrical properties of size and shape.
\ Processes that generate the 1-form $Q$ such that $Q\symbol{94}dQ=\emptyset$
are thermodynamically reversible. \ If $Q\symbol{94}dQ\neq\emptyset,\,$the
process that generates $Q$ is thermodynamically irreversible.

When the Cartan topology is connected, it might be said that all forces are
extendible over the whole of the set, and that these forces are of \ ''long
range''. Conversely when the Cartan topology is disconnected, the ''forces''
cannot be extended indefinitely over the whole \ \ domain of independent
variables, but perhaps only over a single component. \ In this sense, such
forces are said to be of short range, \ as they are confined. \ Note that this
notion of short \ or long range forces does not depend upon geometrical size
or scale. \ The physical idea of short or long range forces is a topological
idea of connectivity, and not a geometrical concept of how far. \vspace{1pt}

In an earlier article, \ these ideas were formulated intuitively in order to
give an explanation of the ''four forces'' of physics. \ The earlier work was
based upon differential geometry, before the construction of the Cartan
topology based upon differential topology as presented herein. [16]\ The
features of the Pfaff sequence were used to establish \ equivalence classes
for known example metric field solutions, $g_{\mu\upsilon},$ to the Einstein
field equations. \ The ideas originally presented upon experience with systems
in differential geometry can now be given credence based upon the construction
differential topology of the disconnected Cartan topology, which will divide
the classes into the long range connected category and the short range
disconnected category. \ The 1-form used to build the Cartan topology was
constructed from the space-time interactions, $A=g_{\mu4}dx^{\mu}\,.$ \ Long
range parity preserving forces due to gravity (Pfaff dimension 1) and
electromagnetism (Pfaff dimension 2) are to be associated with a Cartan
Topology that is connected $(H=A\symbol{94}F=A\symbol{94}dA=0). $ \ Both the
strong force (Pfaff dimension 3) and the weak force (Pfaff dimension 4) are
''short'' range $(H\neq0)$ and are to be associated with a disconnected Cartan
topology. \ The strong force is parity preserving $(K=0)$ and the weak force
is not $(K\neq0).$ \ The fact that the Cartan topology is not necessarily
connect is the topological (not metrical) basis that may be used to
distinguish between short and long range forces.

In much of our physical experience with nature it appears that the
disconnected domains of Pfaff dimension 3 or more are often isolated as
nuclei, while the surrounding connected domains \ of Pfaff dimension 2 or less
appears as fields of \ charged or non-charged molecules and atoms. \ However,
part of the thrust of this article is to demonstrate that such disconnected
topological phenomena are not confined to microscopic systems, but also appear
in a such mundane phenomena as the flow of a turbulent fluid. Physical
examples of the existence of topological torsion (and hence a non-connected
Cartan topology) are given by the experimental appearance of what appear to be
coherent structures in a turbulent fluid flow.

To prove that a turbulent flow must be a consequence of a Cartan topology that
is not connected, consider the following argument: First consider a fluid at
rest and from a global set of unique, synchronous, initial conditions generate
a vector field of flow. Such flows must satisfy the Frobenius complete
integrability theorem, which requires that $A\symbol{94}dA=0$. The Cartan
topology for such systems is connected, and the Pfaff dimension of the domain
is 2 or less. Such domains do not support topological torsion (the Helicity
vanishes). Such globally laminar flows are to be distinguished from flows that
reside on surfaces, but do not admit a unique set of connected sychronizeable
initial conditions. Next consider turbulent flows which, as the anti-thesis of
laminar flows, can not be integrable in the sense of Frobenius; such turbulent
domains support topological torsion $(A\symbol{94}dA\neq0)$, and therefore a
disconnected Cartan topology. The connected components of the disconnected
Cartan topology can be defined as the (topologically)\ coherent structures of
the turbulent flow.

Note that a domain can support a homogeneous topology of one component and
then undergo continuous topological evolution to a domain with some interior
holes. The domain is simply connected in the initial state, and multiply
connected in the final state, but still connected. However, consider the dual
point of view where the originally connected domain consists of a homogeneous
space that becomes separated into multiple components. The evolution to a
topological space of multiple components is not continuous. It follows that
the case of a transition from an initial laminar state ($H=0$) to the
turbulent state ($H\neq0$) is a transition from a connected topology to a
disconnected topology. Therefore the transition to turbulence is NOT
continuous. However, note that the decay of turbulence can be described by a
continuous transformation from a disconnected topology to a connected
topology. Condensation is continuous, gasification is not. It is demonstrated
below that relative to the Cartan topology all C2 differentiable, $\mathbf{V}%
$, acting on C2 p-forms by means of the Lie derivative are continuous. The
conclusion is reached that the transition to turbulence must involve
transformations that are not C2, hence can occur only in the presence of
shocks or tangential discontinuities.

\section{The Cartan Topological Structure}

A topological structure is defined to be enough information to decide whether
a transformation is continuous or not [18]. The classical definition of
continuity depends upon the idea that every open set in the range must have an
inverse image in the domain. This means that topologies must be defined on
both the initial and final state, and that somehow an inverse image must be
defined. Note that the open sets of the final state may be different from the
open sets of the initial state, because the topologies of the two states can
be different.

There is another definition of continuity that is more useful for it depends
only on the transformation and not its inverse explicitly. A transformation is
continuous if and only if the image of the closure of every subset is included
in the closure of the image. This means that the concept of closure and the
concept of transformation must commute for continuous processes. Suppose the
forward image of a \ 1-form $A$ is $Q,$ and the forward image of the set
$F=dA$ is $Z$. Then if the closure,$\,\,A^{c}=A\cup F$ is included in the
closure of$\,\,\,\,Q^{c}=Q\cup dQ$, for all sub-sets, the transformation is
defined to be continuous. The idea of continuity becomes equivalent to the
concept that the forward image $Z$ of the limit points, $dA $, is an element
of the closure of$\,\,Q$ [18]:

\medskip

\begin{quote}
A function $f$ \ that produces an image $f[A]=Q$ is continuous iff for every
subset $A$ of the Cartan topology, $Z=f[dA]\subset Q^{c}=(Q\cup dQ)$.

\medskip
\end{quote}

The Cartan theory of exterior differential systems can now be interpreted as a
topological structure, for every subset of the topology can be tested to see
if the process of closure commutes with the process of transformation. For the
Cartan topology, this emphasis on limit points rather than on open sets is a
more convenient method for determining continuity. A simple evolutionary
process, $X\Rightarrow Y$, is defined by a map $\Phi$. The map, $\Phi$, may be
viewed as a propagator that takes the initial state, $X$, into the final
state, $Y$. For more general physical situations the evolutionary processes
are generated by vector fields of flow, $\mathbf{V}$. The trajectories defined
by the vector fields may be viewed as propagators that carry domains into
ranges in the manner of a convective fluid flow. The evolutionary propagator
of interest to this article is the Lie derivative with respect to a vector
field, $\mathbf{V}$, acting on differential forms, $\Sigma\,$\lbrack19\rbrack.

\medskip

The Lie derivative has a number of interesting and useful properties.

\qquad\qquad1. The Lie derivative does not depend upon a metric or a connection.

\qquad\qquad2. The Lie derivative has a simple action on differential forms
producing a resultant form that is decomposed into a transversal and an exact part:%

\begin{equation}
L_{(\mathbf{V})}\omega=i(V)d\omega+di(V)\omega.
\end{equation}
This formula is known as ''Cartan's magic formula''. \ \ For those vector
fields $V$ which are ''associated'' with the form $\omega,$ such that
$i(V)\omega=0,$ the Lie derivative becomes equivalent to the covariant
derivative of tensor analysis. \ Otherwise the two derivative concepts are distinct.

\qquad\qquad3. The Lie derivative may be used to describe deformations and
topological evolution. Note that the action of the Lie derivative on a 0-form
(scalar function) is the same as the directional or convective derivative of
ordinary calculus,%

\begin{equation}
L_{(\mathbf{V})}\Phi=i(V)d\Phi+di(V)\Phi=i(V)d\Phi+0=\mathbf{V}\cdot grad\Phi.
\end{equation}
It may be demonstrated that the action of the Lie derivative on a 1-form will
generate equations of motion of the hydrodynamic type.

\qquad\qquad4. With respect to vector fields and forms constructed over C2
functions, the Lie derivative commutes with the closure operator. Hence, the
Lie derivative (restricted to C2 functions) generates transformations on
differential forms which are continuous with respect to the Cartan topology.
To prove this claim:

First construct the closure, $\{\Sigma\cup d\Sigma\}$~. Next propagate
$\Sigma\,\ $and $d\Sigma$ by means of the Lie derivative to produce the
decremental forms, say $Q$ and $Z$,%

\begin{equation}
L_{(\mathbf{V})}\Sigma=Q\,\,\,\,\,\,\,\,and\,\,\,\,\,\,L_{(\mathbf{V})}%
d\Sigma=Z.
\end{equation}
Now compute the contributions to the closure of the final state as given by
$\{Q\cup dQ\}$. If $Z=dQ$, then the closure of the initial state is propagated
into the closure of the final state, and the evolutionary process defined by
$\mathbf{V}$ is continuous. \ However,%

\begin{equation}
dQ=dL_{(\mathbf{V})}\Sigma=di(V)d\Sigma+dd(i(V)\Sigma)=di(V)d\Sigma\,\,
\end{equation}
as$\,\,\,\,dd(i(V)\Sigma)=0\,\,\,\,$for C2 functions. But,%

\begin{equation}
Z=L_{(\mathbf{V})}d\Sigma=d(i(V)d\Sigma)+i(V)dd\Sigma=di(V)d\Sigma
\end{equation}
as$\,\,\,i(V)dd\Sigma=0$ for C2 p-forms. \ \ It follows that $Z=dQ$, and
therefore $\mathbf{V}$ generates a continuous evolutionary process relative to
the Cartan topology. $QED$ \ 

Certain special cases arise for those subsets of vector fields that satisfy
the equations, $d(i(\mathbf{V})\Sigma)=0$. In these cases, only the functions
that make up the p-form, ~$\Sigma$, need be C2 differentiable, and the vector
field need only be C1. Such processes will be of interest to symplectic
processes, with Bernoulli-Casimir invariants.

By suitable choice of the 1-form of action it is possible to show that the
action of the Lie derivative on the 1-form of action can generate the Navier
Stokes partial differential equations [20]. The analysis above indicates that
C2 differentiable solutions to the Navier-Stokes equations can not be used to
describe the transition to turbulence. The C2 solutions can, however, describe
the irreversible decay of turbulence to the globally laminar state.

\medskip

\section{\textbf{APPLICATIONS}}

\subsection{Frozen - in Fields, the Master Equation}

A starting point for many discussions of the magnetic dynamo and allied
problems in hydrodynamics starts with what has been called the ''master
equation'' [21],%

\begin{equation}
Curl(\mathbf{V\times B})=\partial\mathbf{B}/\partial t.
\end{equation}

Using the Cartan methods it may be shown that this equation is equivalent to
the constraint of ''uniform'' continuity relative to the Cartan topology.
Moreover, it is easy to show these constraints\ generate symplectic processes
which include Hamiltonian evolutionary systems, such as Euler flows, as well
as a number of other evolutionary processes which are continuous, but not
homeomorphic. In addition a criteria can be formulated to develop an extension
of the ''helicity'' conservation law to a more general setting.

The proof of these results produces a nice exercise in use of the Cartan
theory. Consider a 1-form $A$ that satisfies the exterior differential system%

\begin{equation}
F-dA=0,
\end{equation}
where $A$ is a 1-form of Action, with twice differentiable coefficients
(potentials proportional to momenta) which induce a 2-form, $F,$ of
electromagnetic intensities ($\mathbf{E}$ and $\mathbf{B\,,\,\,\,}$related to
forces$)$. \ The exterior \ differential system is a topological constraint
that in effect defines field intensities in terms of the potentials. On a four
dimensional space-time of independent variables, $(x,y,z,t)$ the 1-form of
Action (representing the postulate of potentials) can be written in the form%

\begin{equation}
A=\Sigma_{k=1}^{3}A_{k}(x,y,z,t)dx^{k}-\phi(x,y,z,t)dt=\mathbf{A\circ
}d\mathbf{r-}\phi dt.
\end{equation}

Subject to the constraint of the exterior differential system, the 2-form of
field intensities, $F,$ becomes:%

\begin{align}
F  &  =dA=\{\partial A_{k}/\partial x^{j}-\partial A_{j}/\partial
x^{k}\}dx^{j}\symbol{94}dx^{k}=F_{jk}dx^{j}\symbol{94}dx^{k}\\
&  =\mathbf{B}_{z}dx\symbol{94}dy+\mathbf{B}_{x}dy\symbol{94}dz+\mathbf{B}%
_{y}dz\symbol{94}dx+\mathbf{E}_{x}dx\symbol{94}dt+\mathbf{E}_{y}%
dy\symbol{94}dt+\mathbf{E}_{z}dz\symbol{94}dt.\nonumber
\end{align}
where in usual engineering notation,%

\begin{equation}
\mathbf{E}=-\partial\mathbf{A}/\partial t-grad\phi\mathbf{,\,\,\,\;\;\;\;\;B=}%
curl\,\,\mathbf{A}\equiv\partial A_{k}/\partial x^{j}-\partial A_{j}/\partial
x^{k}.
\end{equation}

The closure of the exterior differential system, $dF=0,\,$\ vanishes for C2
differentiable p-forms, to yield%

\begin{equation}
dF=ddA=\{curl\;\mathbf{E}+\partial\mathbf{B}/\partial t\}_{x}dy\symbol{94}%
dz\symbol{94}dt-..+..-div\,\mathbf{B}dx\symbol{94}dy\symbol{94}dz\}\Rightarrow
0.
\end{equation}
Equating to zero all four coefficients leads to the Maxwell-Faraday equations,%

\begin{equation}
\{curl\;\mathbf{E}+\partial\mathbf{B}/\partial
t=0,\,\,\,\,\,\,\,\,\,\,div\,\mathbf{B}=0\}.
\end{equation}

The component functions ($\mathbf{E}$ and $\mathbf{B})\,$of the 2-form, $F,$
transform as covariant tensor of rank 2. \ \ The topological constraint that
$F$\ is exact, implies that the domain of support for the field intensities
cannot be compact without boundary, unless the Euler characteristic vanishes.
These facts distinguish classical electromagnetism from Yang-Mills field
theories. \ Moreover, the fact that $F$ is subsumed to be exact and C1
differentiable excludes the concept of magnetic monopoles from classical
electromagnetic theory on topological grounds.

Now search for all vector fields that leave the 2-form $F$ an absolute
invariant of the flow; that is, search for all vectors that satisfy Cartan's
magic formula%

\begin{equation}
L_{(\mathbf{V})}F=i(V)dF+di(V)F=0+di(V)F=0.
\end{equation}
\ For C2 functions, the term involving $dF$ vanishes, leaving the expression,%

\begin{align}
L_{(\mathbf{V})}F  &  =di(V)F\\
&  =d\{(\mathbf{E}+\mathbf{V\times B})\cdot\,d\mathbf{r-(E\cdot V})dt\}\\
&  =\{curl(\mathbf{E}+\mathbf{V\times B})\}_{z}dy\symbol{94}dz...\\
&  +\{\partial(\mathbf{E}+\mathbf{V\times B)}/\partial t+grad(\mathbf{E\cdot
V})\cdot\,d\mathbf{r}\symbol{94}dt\\
&  =0.
\end{align}
Setting the first three factors to zero yields%

\begin{equation}
curl\mathbf{(E+V\times B})=0
\end{equation}

But for C2 functions, $curl\mathbf{E}=-\partial\mathbf{B}/\partial t$, and
when this expression is substituted into the above equation, the ''master
equation given by \ the first equation results. Now recall that $dF$ generates
the limit points of $A$, and if $F=dA$ is a flow invariant, then all limit
points are flow invariants relative to the Cartan topology. This result
implies that the vector fields, $\mathbf{V}$, that satisfy the constraints of
the ''master equation'' are uniformly continuous evolutionary processes, the
limit points, $F=dA,$ of the 1-form $\ A$ are flow invariants, and the lines
of vorticity are ''frozen-in'' the flow. \ Non-uniform continuity would imply
that the limit points are not invariants of the process, but that the closure
of the limit points of the target range include the vanishes limit points of
the initial domain. \ Such processes would correspond to a folding of the
''lines'' \ of vorticity, which preserve the limit points, but not their
sequential order.

A second criteria for limit point invariance is given by the equation,%

\begin{equation}
\{\partial(\mathbf{E}+\mathbf{V\times B)}/\partial t+grad(\mathbf{E\cdot
V})\}=0.
\end{equation}
The formula indicates that \ limit point invariance can occur in the presence
of dissipation, $\mathbf{E\cdot V}\neq0.$

The criteria for frozen-in fields is established as a constraint on the
admissable vector fields, $\,di(V)dA=di(V)F=0.$ \ The solution vector fields,
$V,$ subject to this constraint can be put into three global categories:\smallskip

\qquad1. \ Extremal (Hamiltonian) \ \ \ $\qquad\qquad\qquad\ \ \ \ \ \ \ i(V)F=0.$

\qquad2. \ Bernoulli-Casimir (Hamiltonian) \ $\qquad\qquad
\ \,\,\,i(V)F=d\Theta.$

\qquad3. \ Symplectic\qquad\qquad\qquad\qquad\qquad\qquad
\ \ \ \ $\ \ \ i(V)F=d\Phi+\gamma_{harmonic}\smallskip$

The first category can exist only on domains of support of $F$ which are\ of
odd Pfaff dimension, but then the solution vector is unique to within a
factor. \ In the other categories, the solution vector need not be unique.
\ Vector fields that satisfy the equation for uniform continuity are said to
be symplectic relative to the 1-form, $A.$\ \ Vector fields that belong to
categories 1 and 2 have a Hamiltonian representation. \ Vector fields that
belong to category 1, are said to be ''extremal'' relative to the 1-form, $A. $

\subsection{Euler flows and Hamiltonian systems.}

\qquad In 1922 Cartan established the idea that the necessary and sufficient
conditions for a system to admit a unique Hamiltonian representation for its
evolution, $\mathbf{V}$, is given by the category 1 constraint,%

\begin{equation}
W=i(V)dA=i(V)F=0.
\end{equation}
It is apparent that this extremal condition is more stringent than that given
above for uniform continuity, $di(V)F=0$. \ Such extremal vector fields are
independent of parameterization. \ That is, for extremal processes, $i(\rho
V)dA=0$ if $\,\,i(V)dA=0,\,\,$for any function, $\rho.$ \ \ Extremal vector
fields do not exist on domains where the Pfaff dimension of the Cartan 1-form
is even. \ In classical mechanics, the 1-form $W$ is defined as the 1-form of
Virtual Work, and the Cartan constraint is typical of problems in the
variational calculus where it is presumed that the Virtual \ Work vanishes.\ \ 

As an example,consider a 1-form of Action defined as%

\begin{equation}
A=\mathbf{v}~d\mathbf{r}-(\mathbf{v~\cdot\,v}/2+\Psi)dt,
\end{equation}
where $d\Psi=dP/\rho$. \ Application of the extremal constraint yields the
resulting necessary system of partial differential equations is given by known
as the Euler equations of hydrodynamics.%

\begin{equation}
\partial\mathbf{v}/\partial t+grad(\mathbf{v\cdot v}/2)-\mathbf{v\times
w}=-gradP/\rho,
\end{equation}

\vspace{1pt}It also follows that the Master equation is valid, with the only
difference being that $curl\mathbf{v}$ is defined as $\mathbf{\omega}$, the
vorticity of the hydrodynamic flow. The master equation becomes,%

\begin{equation}
curl(\mathbf{v\times\omega})=\partial\mathbf{\omega}/\partial t,
\end{equation}
and is to be recognized as Helmoltz' equation for the conservation of
vorticity. In the hydrodynamic sense, conservation of vorticity implies
uniform continuity. \ In other words, the Eulerian flow is not only
Hamiltonian, it is also uniformly continuous, and satisfies the master
equation and the conservation of vorticity constraints. In addition, it may be
demonstrated that such systems are at most of Pfaff dimension 3, and admit a
relative integral invariant which generalizes the hydrodynamic concept of
invariant helicity. In the electromagnetic topology, the Hamiltonian
constraint is equivalent to the statement that the Lorentz force vanishes, a
condition that has been used to define the ''ideal'' plasma or ''force-free''
plasma state.

\subsection{Conservation of Topological Torsion}

A slightly more general class of evolutionary processes (flows) is given by
the constraints which are gauge equivalent to the Hamiltonian extremal case; a
search is made for those flows that satisfy the (non-exrtremal, but
Hamiltonian) constraint:%

\begin{equation}
i(\rho V)dA=i(\rho V)F=dW.
\end{equation}
Such flows admit two topological invariants of the relative integral invariant
form. \ The first integral invariant is 1-dimensional:%

\begin{align}
L_{(\rho\mathbf{V})}\oint_{1d\_closed}A  &  =\oint_{1d\_closed}i(\rho
V)dA+di(\rho V)A=\\
\oint_{1d\_closed}dW+di(\rho V)A  &  =\oint_{1d\_closed}d\{W-i(\rho
V)A\}\Rightarrow0,
\end{align}
expressing the relative integral invariance of circulation (Kelvin's theorem).
\ The second integral invariant is 3-dimensional:%

\begin{equation}
L_{(\rho\mathbf{V})}\oint_{3d\_closed}A\symbol{94}dA=\oint_{3d\_closed}%
d\{W-i(\rho V)A\}\symbol{94}dA\Rightarrow0,
\end{equation}
a result expressing the generalization of the law which in hydrodynamics is
called the conservation of Helicity. \ The integrations are over closed 1 and
3 dimensional domains. \ These closed integration domains can be either cycles
or boundaries. \ For exampled the 1-dimensional closed curve in the punctured
disc that encircles the central hole is a cycle but not a boundary. \ As the
integrands are exact differentials, the closed integrals vanish.

Note that on the domain $\{x,y,z,t\}$, the 3-form of topological torsion,
$A\symbol{94}dA$, has the general representation with coefficients, $Z_{\mu
\nu\sigma}$, that transform as a covariant tensor field of third rank. \ On a
4 dimensional space, the components of $A\symbol{94}dA$ are proportional to a
contravariant tensor density of rank 1, whose four components have a vector
part defined as, $\mathbf{T}$, the torsion (pseudo) current, and a (pseudo)
density part, $h$. The 3-form $A\symbol{94}dA$ is not an impair form
(density). \ In electromagnetic engineering language, the general formula for
the torsion 3-form has a component expression given by:%

\begin{equation}
T=[\mathbf{T},h]=[\mathbf{E\times A}+\phi\mathbf{B},\mathbf{A\cdot B}].
\end{equation}
For the constraints of an Eulerian flow, the 4 components of the Torsion three
form reduce to%

\begin{equation}
T=[\mathbf{T},h]=[(\mathbf{v\cdot\omega)v}-(\mathbf{v\cdot~v}/2+\Psi
)\mathbf{\omega,v\cdot\omega].}%
\end{equation}

Recall that the closed integration domain used to evaluate the relative
integral invariant is \textit{not} necessarily restricted to a spatial volume
integral with a boundary upon which the normal component of \ $\mathbf{v}$
vanishes. Also note that the helicity density of hydrodynamic fame is the
fourth component, $h=\mathbf{v\cdot\omega,}$ of a contravariant vector
density, equivalent to a covariant tensor of third rank. Care must be used in
its transformation with respect to diffeomorphisms, such as the Galilean
transformation. Furthermore, for the constraints of an Eulerian flow (an
extremal field) described above, the topological parity 4-form vanishes
globally, such that there exists a pointwise conservation law of the 3-form,
equivalent to the expression,%

\begin{equation}
div_{3}\mathbf{T}+\partial h/\partial t=0.
\end{equation}

\subsection{Topological Invariants and Period Integrals}

Besides the invariant structures considered above, the Cartan methods may be
used to generate other sets of topological invariants. Realize that over a
domain of Pfaff dimension $n\leq N$, the Cartan criteria admits a submersive
map to be made from N to a space of minimal dimension n. The map may be viewed
as a vector field of functional components,%

\[
\lbrack V^{x}(x,y,z..),V^{y}(x,y,z..),V^{z}(x,y,z..),...],
\]
of dimension n, and will have a representation in the projective geometry of
n+1 homogeneous coordinates. The n+1 component will be generated by a function
$\lambda,\,$related to the Holder norm,%

\begin{equation}
\vspace{1pt}\rho=1/\lambda=1/\{a(V^{x})^{p}+b(V^{y})^{p}+c(V^{z}%
)^{p}+.....\}^{n/p}.
\end{equation}
For any vector field, construct the n dimensional volume element,%

\begin{equation}
\Omega=\rho(V)\,\,dV^{x}\symbol{94}dV^{y}\symbol{94}dV^{z}...
\end{equation}
and the n-1 form density (current) $J\,\ $as:%

\begin{align}
J  &  =\,i(V^{x},V^{y},V^{z},...)\Omega=\nonumber\\
&  \rho\{V^{x}\,\,dV^{y}\symbol{94}dV^{z}...\,\,-V^{y}\,\,dV^{x}%
\symbol{94}dV^{z}...+V^{z}\,\,dV^{x}\symbol{94}dV^{y}...-...\}\ \ \ \ \ \ .
\end{align}
It is remarkable that the current $J$ so defined has a vanishing exterior
derivative, independent of the value of p for a given n, and for all values of
the constants, plus or minus a,b,c...). \ All such currents define a
''conservation law''. As the map defining the components of the vector field
in terms of the base \{x,y,z..\} is presumed to be differentiable, then the
n-1 form, $J,$ has a well defined pull back on the base space (almost every
where), and its exterior derivative on the base space also vanishes everywhere
mod the defects. \ That is, the form $J$ is locally exact.

In the expression for $\lambda,$ the factors $\{a,b,c,d...\}$ are arbitrary
constants of either sign. \ \ The most familiar format is when $p=2,$ and then
the function $\lambda$ has a null set which is a conic. For positive isotropic
signature, the only defect is the origin in the space defined by the
functions, $V.$ \ \ The construction produces the algebraic dual or adjoint
vector field from the functional components of the original vector field with
integrating factors $\rho=1/\lambda$ that create conservation laws for
physical systems. \ The integrals of these closed currents when integrated
over closed N-1 dimensional chains form deformation invariants, with respect
to any evolutionary process that can be described by a vector field,
for\vspace{1pt}
\begin{equation}
L_{(\rho\mathbf{V})}\oint_{n-1}J=%
{\textstyle\oint_{n-1}}
i(\rho\mathbf{V})dJ+%
{\textstyle\oint_{n-1}}
d(i(\rho\mathbf{V}))J)=0+0=0
\end{equation}
These integral objects appear as ''topological coherent'' structures, which
may have defects or anomalous sources, when the integrating factor $1/\lambda$
is not defined.

The compliment to the zero sets of the function $\lambda$ determine the domain
of support associated with the specified vector field. The closed n-1 form,
$J$, that satisfies the conservation law, $dJ=0$, has integrals over closed
domains that have rational fraction ratios. \ As this n-1 current is closed
globally, it may be deduced on a connected local domain from a n-2 form, $G$.
In every case $J$ has a well defined pull-back to the base variety, x,y,z,t.
Note that the n functions $[V^{x}\,(x,y,z..),V^{y}(x,y,z..),V^{z}%
(x,y,z..),...]$ represent the minimum number of Clebsch variables that are
equivalent to the original action, $A$, over the domain of support. As each of
these integrals is intrinsically closed, the Lie derivative with respect to
any C2 vector field, $\mathbf{V}$, is a perfect differential, such that (when
integrated over closed domains that are p-1 boundaries) the evolutionary
variation of these closed integrals vanishes. These n-1 integrals are relative
integral invariants for any C2 evolutionary processes, or flows. The values of
the integrals are zero if the closed integration domains are boundaries, or
completely enclose a simply connected region. If the closed integration
domains encircle the zeros of the function $\lambda$, then the values of the
integrals are proportional to the integers; i.e., their ratios are rational.
Note that each signature must be investigated. For the elliptic (positive
definite) signature, the singular points are the stagnation points, and the
domain of support excludes those singularities. For the hyperbolic signatures,
the domain of support excludes the hyperbolic singularities of lower
dimension, such as the light cone. Further note that a given vector field may
not generate real domains of support for all possible signatures of the
quadratic form, $\lambda$.

\medskip

\subsection{The Flux or Circulation Integral 1-form}

\qquad For the Cartan topology constructed from a fundamental 1-form of Action
and a fundamental N-1 form of Current, several period integrals of closed
forms integrated over closed chains appear in a natural manner. \ In
particular on an N=4 dimensional domain, the four period integrals of most
interest are the period integrals of flux (circulation), charge, spin and
torsion \lbrack9\rbrack. \ The fundamental period integral over a closed
1-form will be defined as the ''Circulation'' or ''flux'' integral. \ When the
Pfaff dimension is 2, there exists a submersive map to two dimensions, and the
vector fields on this domain will have two irreducible components, say
\lbrack$\Phi(x,y,z,t),\Psi(x,y,z,t)\rbrack.$ \ Following the procedure of the
preceding section, construct the 2-dimensional volume element defined as
$\Omega=\rho d\Phi\symbol{94}d\Psi,$ and the $n-1=2-1=$ 1 form $A=(\Phi
d\Psi-\Psi d\Phi)/\{\pm a\Phi^{p}\pm b\Psi^{p}\}^{2/p}.$ \ The exterior
derivative of such a 1-form is exactly zero for all point sets that exclude
the null set of the denominator. \ The classic choice is for $\ p=2,$and
$a=1,\,b=1,$ (+,+) signature. The closed integrals of these closed 1-forms
then can be expressed as%

\begin{equation}
\operatorname{Ci}rculation\,\,\,\Gamma=%
{\textstyle\oint_{1}}
A=%
{\textstyle\oint_{1}}
(\Phi d\Psi-\Psi d\Phi)/\{\Phi^{2}+\Psi^{2}\}
\end{equation}
By substituting the functional forms in terms of (x,y,z,t) the circulation
integral can be written in terms of functions on (x,y,z,t) and their
differentials, $\{dx,\,dy,\,dz,\,dt...\}$ \ \ \ 

As an example, suppose that the domain is three dimensional, N=3. \ Then the
zero sets of $\Phi(x,y,z)=0$ and $\Psi(x,y,z)=0,$ represent two 2 dimensional
surfaces which may or may not have one or more lines of \ intersection. \ If
the surfaces intersect, then%

\begin{equation}
Intersection=d\Phi\symbol{94}d\Psi\neq0.
\end{equation}
If the closed integration paths cannot be contracted to a point, because they
encircle these lines of intersection, the values of the integrals have
rational ratios depending on how many lines are encircled and how many times
the integration path encircles a line. \ The lines of intersection must have
zero divergence (and therefore must stop or start on boundary points, or are
closed on themselves). \ Otherwise the \ integration chains can be deformed
and then contracted to a point. \ The classic example is given by the 1-form,
$A=(ydx-xdy)/(+x^{2}+y^{2})$ in three dimensions. \ For integration contours
that encircle the z axis, the value of $\Gamma=%
{\textstyle\oint_{1}}
A=2\pi.$ \ In hydrodynamics, this vector field is called a potential
''vortex'', even though the vorticity $\mathbf{\omega=}curl\mathbf{v}%
=0\mathbf{.}$ \ Stokes theorem does not apply as the closed integration chain
is a cycle that is not a boundary.

\qquad An interesting application of the circulation integral is given when
there exists a map to the complex domain. \ Then $\Psi\Rightarrow\Phi^{\ast}$
and the circulation integral has the form of the integral of the probability
current in standard quantum mechanics.%

\begin{equation}
Period=%
{\textstyle\oint_{1}}
(\Phi d\Phi^{\ast}-\Phi^{\ast}d\Phi)/\{\Phi\cdot\Phi^{\ast}\}.
\end{equation}

\medskip

\subsection{The Gauss Linking or Charge Integral 2-form}

\qquad Many different options exist for construction of these invariant
topological structures from closed p-forms. \ The idea is to find a
formulation for a closed form on a domain, and then to specify a closed and
compatible integration chain. \ The integration chain need not be a boundary,
but only a closed cycle. \ For example, from the components of the specified
vector, $A_{\mu}$ , the Jacobian matrix, $\left[  \partial A_{\mu}/\partial
x^{\nu}\right]  $ can be constructed. The rows or columns of the matrix of
cofactors of the Jacobian (the adjoint matrix) forms a set of vector fields
that have zero divergence \lbrack21\rbrack, and therefore these vectors could
be used to construct relative integral invariants. In every case there exists
an algebraic construction which produces a vector that is divergence free and
whose line of action is uniquely related to original vector that was used to
construct the Cartan topology. That vector may be constructed by multiplying
the original vector $A_{\mu}$ by the matrix of cofactors and then dividing by
the function $\lambda$ defined above. The construction replicates the previous
procedure. As an application for n = 3, p=2, consider the vector that
represents the difference between two space curves,$\,\,\,\mathbf{z=R}%
_{2}\mathbf{-R}_{1}$. Then compute the two form $G(z)$ from the ''volume''
element $\Omega=dz^{1}\symbol{94}dz^{2}\symbol{94}dz^{3}/\lambda,$ to give%

\begin{equation}
\vspace{1pt}G_{n=3}=\{z^{1}dz^{2}\symbol{94}dz^{3}-z^{2}dz^{3}\symbol{94}%
dz^{1}+z^{3}dz^{1}\symbol{94}dz^{2}\}/\lambda
\end{equation}
where%

\begin{equation}
\lambda=(\pm(z^{1})^{2}\pm(z^{2})^{2}\pm(z^{3})^{2})^{3/2}.
\end{equation}

Next assert that the displacements of interest are constrained by two
parametric curves given by%

\begin{equation}
d\mathbf{R}_{1}=\mathbf{V}_{1}dt\,\,\,\,\,\,\,\,and\,\,\,\,\,\,\,\,d\mathbf{R}%
_{2}=\mathbf{V}_{2}dt^{\prime},
\end{equation}
where the parameters $dt$ and $dt\prime$ are not functionally related (which
would imply that $dt\symbol{94}dt\prime=0$).

\qquad It is important to realize that kinematic constraints are topological
constraints that refine the Cartan topology, \ a topology based solely upon
the specified 1-form of action, $A$. \ \ From a physical point of view, these
constraints can be interpreted as constraints of null fluctuations and in
certain circumstances can be associated physically with the limit of zero
temperature. To demonstrate the utility of such constraints, substitute these
differential expressions into the expression for the 2-form $G$ of ''current''
in N=3 dimensions, and carry out the exterior products, using $dt\symbol{94}%
dt^{\prime}\neq0,$ $but\ dt\symbol{94}dt=0$ and $dt\prime\symbol{94}%
dt\prime=0.$ \ The result is the vector triple product representation for the
Gauss integral,%

\begin{equation}
\vspace{1pt}Q=%
{\textstyle\oint_{2}}
G=%
{\textstyle\oint_{2}}
\{\mathbf{z\circ V}_{1}\times\mathbf{V}_{2}\}dt\symbol{94}dt^{\prime
}/(\mathbf{R}_{1}\circ\mathbf{R}_{1}-2\mathbf{R}_{1}\circ\mathbf{R}%
_{21}+\mathbf{R}_{2}\circ\mathbf{R}_{2})^{3/2}.
\end{equation}

The integration domain is the closed ''2-dimensional area'' formed by the
displacements along the non-intersecting curves defined by the two distinct
parameters, $dt$, and $dt^{\prime}.$ \ This double integral is to be
recognized as the Gauss linking integral of Knot Theory \lbrack7\rbrack.
\ \ (Without the kinematic substitutions, it may also be interpreted as the
charge integral of electromagnetic theory.) When integrations are computed
along closed curves whose tangent vectors are $\mathbf{V}_{1}$ and
$\mathbf{V}_{2}$, then the integer values of the closed integral may be
interpreted as how many times the two curves are linked. Note that the same
integer result is obtained when the vector $\mathbf{z}$ is interpreted as the
sum of the two vectors, $\,\,\mathbf{z=R}_{2}+\mathbf{R}_{1}$, although the
values of the integrals have different scales. \ 

The constraint that $dt\symbol{94}dt^{\prime}\neq0$ implies that the
''motion'' along the curve generated by $\mathbf{R}_{1}$ is independent of the
''motion'' along the curve generated by $\mathbf{R}_{2}.$ \ If the curve
generated by $\mathbf{R}_{1}$ is a conic in the $xy$ plane and the curve
generated by $\mathbf{R}_{2}$ is a conic in the $xz$ plane, then the surface
swept out by the vector $\mathbf{z}$ is a Dupin cyclide. \ Such surfaces have
application to the propagation of waves in electromagnetic systems.

From another point of view, consider the ruled surface [22]\ defined by the
vector field of two parameters,%

\begin{equation}
\vspace{1pt}\,\,\mathbf{z(}\mu,t\mathbf{)=R}(t)\pm\mu\mathbf{V}(t).
\end{equation}
Vector fields of this type are primitive types of ''strings'' for fixed values
of the parameter, $t$, and string parameter, $\mu.\,\,\,$ Direct substitution
of the physical constraints, $d\mathbf{R}-\mathbf{V}dt=0$, and $d(\mathbf{V)}%
-\mathbf{A}dt=0$ leads to the topological Gauss integral,\vspace{1pt}%

\begin{align}
\vspace{1pt}\vspace{1pt}Q  &  =%
{\textstyle\oint_{2}}
G=%
{\textstyle\oint_{2}}
\{\mathbf{R\circ}\mu\mathbf{V}\times\mathbf{A}\}/\lambda\mathbf{=}\nonumber\\
&
{\textstyle\oint_{2}}
\{\mathbf{A\circ R\times}\mu\mathbf{V}\}dt\symbol{94}d\mu/(\mathbf{R}%
\circ\mathbf{R}\pm2\mu\mathbf{R}\circ\mathbf{V}+\mu\mathbf{V}\circ
\mu\mathbf{V})^{3/2}.
\end{align}

\qquad It is apparent that the interaction of the ''angular'' momentum,
$\mathbf{L}=\mathbf{R}\times\mu\mathbf{V}$, and the acceleration, $\mathbf{A}%
$, produces a topological invariant whose values are ''quantized'' ( in the
sense that the ratios of the integrals are rational). Note that for the
classical central field problem where the force (acceleration) and the angular
momentum are orthogonal, the orbits are in a plane and the Gauss--linking
number is zero. Further note that the triple vector product of the integrand
is proportional to the Frenet torsion of the orbit. An orbit that is planar
has Frenet torsion zero everywhere. The Gauss linking integral is a special
case of the Gauss two dimensional period integral of electromagnetic theory
when the integration domains can be factored into independent products,
$dt\symbol{94}dt\prime\neq0.$

\qquad

\subsection{Chaos and the Unknot}

Much interest of late has been shown in knot theory and its application to an
understanding of the trajectories of dynamical systems. The conjecture is that
somehow an understanding of knot theory will give a better understanding of
chaos. Counter intuitively is the idea that chaos is to be related to the
unknot. Of particular interest will be those cases where lines of vorticity
have an oscillatory Frenet torsion with a period equal to 2/3 of the
fundamental period of closure. The topological Gauss integral will average to
zero for such systems; but these systems can be created by continuous
deformations of folding and twisting a closed loop of vorticity, producing a
period 3 system which is known to be related to chaos [23]. \ In the
undeformed circular state, tubular neighborhoods guided by the vortex lines
can continuously evolve into domains without stagnation points or tangential
singularities, or knots, or twists. However, when the closed vortex line is in
the deformed period 3 configuration, tangential (hyperbolic) singularities are
created by the flow lines of the velocity field, and the evolution becomes
highly convoluted and chaotic. See Figure 1.

These topological features may be demonstrated visually by taking a long strip
of paper and wrapping the strip three times around your fingers. Close the
strip by going under one strand and over the next before pasting together. The
strip is of obvious period three. Now slide the closed strip from the fingers
and note that it can be deformed 9continuously into a cylindrical strip
without twists or knots (Spin 0). If the same procedure is used, except that a
double over or a double under crossing is used before pasting the strip ends
together, the resulting closed loop will have a continuously irreducible
$4\pi$ twist (Spin 2). Both the Spin 2 and the Spin 0 strips have a zero Euler
characteristic. However, the Spin 2 strip can be continuously deformed into a
Klein bottle, or a double lapped Mobius band, and is not homeomorphic to the
spin zero strip [24].

If a model of the Spin0 and Spin 2 systems (deformed to their period 3
configurations) is made from a copper tube, and if flexible bands are created
to link any pair of neighboring tubular strands, then it is readily observed
that the paired domain twists and folds as it is propagated unidirectionally
along the vortex lines. For the spin 2 system the flexible bands will return
to their original state in 3 revolutions. However, the paired domain continues
to twist and fold, becoming ever more complicated as it follows the evolution
around the Spin 0 configuration. The folded spin 0 system has chaotic
neighborhoods. This result indicates that the source of chaos in dynamical
systems may be due to the unknot, and not the knot! The Cartan theory thereby
predicts that the source of chaos in turbulent systems does not require a
discontinuous cut and connect process, but may be induced by vortex lines that
continuously evolve by twisting and folding into a closed, spin 0, period
three configuration.

\medskip

\subsection{The Torsion 3-form and the Braid integral}

For n = 4 the same procedures used above can be used to produce a period
integral over a closed 3-dimensional domain. In fact, the same vector field
that is used to define the Cartan 1-form of Action may be used to construct a
dual N-1 form that is closed. The algorithm is to substitute for the functions
$\lbrack V^{x}(x,y,z..),V^{y}(x,y,z..),V^{z}(x,y,z..),...\rbrack$ the
functions $\lbrack A_{x},A_{y},A_{z}...\rbrack$,~ that make up the covariant
1-form of Action.. This construction is equivalent to constructing the
Jacobian matrix of the original vector field on the N-dimensional velocity
space, computing its cofactor matrix, multiplying the original vector by the
cofactor matrix, and then dividing by the quadratic form, $\lambda$. When
these operations are completed, functional substitution will lead to an
conserved axial vector current density on (x,y,z,t). Another form of the
topological integral invariant is constructed in the following way. First, for
the classic Cartan action, $A=P_{k}dx^{k}-Edt/c$, construct the N-volume,
$\Omega=-dP_{x}\symbol{94}dP_{y}\symbol{94}dP_{z}\symbol{94}dE/c.$ \ Next
contract $\Omega$ with the vector, $(Px,Py,Pz,-E/c)$, and then divide by
$\lambda=\{\pm P\circ P\pm(E/c)^{2}\}^{2}.$ For sake of simplicity, assume
that $E/c$ is a constant such the $dE=0$. Then the closed 3-form or current
becomes equivalent to%

\begin{equation}
J=(E/c)dP_{x}\symbol{94}dP_{y}\symbol{94}dP_{z}/\lambda
\,\,\,\,\,\,\,\,\,\,\,\,\,\,\,\,\,\,\,\,\,\,\,\,\,\,\,\,\,with\,\,\,\,\,\,\,\,\,\,\,\,\,dJ=0
\end{equation}

Now invoke the same Cartan trick of individual parametrization as uses above.
Consider a total momentum vector composed of three individual vector
components, $\mathbf{P=p}_{\mathbf{1}}\mathbf{+p}_{\mathbf{2}}\mathbf{+p}%
_{\mathbf{3}}.$ Assume that the Cartan topology is constrained in such a way
that for each vector component a Newtonian kinematic law of parametrization is
maintained such that%

\begin{equation}
\vspace{1pt}d\mathbf{p}_{1}\mathbf{-f}_{1}dt=0,\,\,\,\,\,\vspace
{1pt}d\mathbf{p}_{2}\mathbf{-f}_{2}dt^{\prime}=0,\,\,\,\,\,\,\,\vspace
{1pt}d\mathbf{p}_{3}\mathbf{-f}_{3}dt^{\prime\prime}=0.
\end{equation}

Also note that $dt\symbol{94}dt^{\prime}\symbol{94}dt^{\prime\prime}\neq0;$
that is, the parameters used in the Newtonian kinematic descriptions are not
sychronizeable. If they were functionally related the value of J must be zero.
Substitute these expressions into the equation for the closed current $J$ and
integrate over a closed 3 dimensional chain to yield a triple Braid integral,%

\begin{align}
Braid  &  =%
{\textstyle\oint_{3}}
J=%
{\textstyle\oint_{3}}
(E/c)dP_{x}\symbol{94}dP_{y}\symbol{94}dP_{z}/\lambda\,\,\nonumber\\
&  =%
{\textstyle\oint_{3}}
(E/c)\{\,\mathbf{f}_{1}\circ(\mathbf{f}_{2}\times\mathbf{f}_{3}%
)\}\,dt\symbol{94}dt^{\prime}\symbol{94}dt^{\prime\prime}/\,\{\pm
\mathbf{P}\circ\mathbf{P}\pm(E/c)^{2}\}^{2}\,\,\,
\end{align}

The integrations are now over three closed curves whose tangents are the
Newtonian forces, $\mathbf{f}$, on three ''particles''. Where in the two
dimensional Gauss integral, of the previous section, the evaluation was along
the closed curves of two ~particles that formed the ends of a string, in this
case the integrations are along the closed trajectories of three ''particles''
which form the vertices of a triangle. In every case, the trajectories are the
trajectories of a system of limit points.

The idea that three ''lines'' are used to form the integral (whose values form
rational ratios) is the reason that this topological integral in the format
given above is defined as the braid integral. Of course the three form of
topological torsion is a variant of the braid integral, but applies to those
topologies where the system is not reducible to three factors $dt,dt\prime$
and $dt"$~ (such systems are said to have torsion cycles). An example of a
period 3 braid with Braid integral zero (chaotic) and Braid integral 2
(non-chaotic) is given in Figure 1

\vspace{1pt} The equivalent to this Figure, and the fact that there are two
distinct period 3 configurations, one chaotic and one non-chaotic, was brought
to the attention of the present authors during a stimulating lecture given by
J. Los at the August, 1991, Pedagogical Workshop on Topological Fluid
Mechanics held at the Institute for Theoretical Physics, Santa Barbara UCSB.

It is to be noted that the 3-form of topological torsion is related to the
braid integral, a three dimensional thing in four dimensions, and not the
Gauss linkage integral, which is a two dimensional thing in three dimensions.
\ The literature of helicity is sometimes confused on this point, and often
attempts to relate the helicity integral to the linkage integral.

\medskip

\subsection{Navier Stokes flows and Pfaff Dimension 4}

\qquad As a last example consider a system where the strong kinematic
(topological) constraint $d\mathbf{x}-\mathbf{V}dt=0$ is not imposed apriori.
\ In other words, the admissable evolutionary processes, $\mathbf{V} $, may
have anholonomic fluctuations about kinematic perfection.%

\begin{equation}
\Delta\mathbf{x}=d\mathbf{x}-\mathbf{V}dt\neq0\
\end{equation}
The physical system will be built on the Cartan topology of the 1-form, $A$,
given previously for the Euler flow. However, the Cartan topology will be
constrained, not by the Hamiltonian conditions required to generate an
extremal system ( which is free of kinematic fluctuations), but by a more
relaxed set of conditions that permit finite kinematic fluctuations,
$d\mathbf{x}-\mathbf{V}dt\neq0$. As it is known that $i(V)dA$ must be
transversal to the vector field, $V$, it follows that a weaker topological
constraint might exist in the form,%

\begin{equation}
i(V)dA=f_{k}(dx^{k}-V^{k}dt)+d\theta,
\end{equation}
where the functions $\theta$ are Bernoulli-Casimir first integrals in the
sense that $i(V)d\theta=0.$

When\thinspace\thinspace\thinspace\ $f_{k}=0$, these fluctuation constraints
reduce to the more stringent Hamiltonian conditions for an extremal flow, or
in the case where $d\theta\neq0,$ to the Bernoulli-Casimir symplectic
conditions. If is assumed that%

\begin{equation}
f_{k}=v(curlcurl\,\mathbf{V})\,_{k},
\end{equation}
it follows that the expression given above, $i(v)dA=f_{k}(dx^{k}-V^{k}dt),$ is
exactly equivalent to the Navier-Stokes partial differential system [25]\ for
an incompressible viscous flow on the variety $x,y,z,t$.%

\begin{equation}
\vspace{1pt}\{\partial\mathbf{V}/\partial t+grad(\mathbf{V\circ V}%
/2)-\mathbf{V}\times curl\mathbf{V\}=}\{\nu\nabla^{2}\mathbf{V\}-}grad\,P/\rho
\end{equation}

These relaxed topological constraints, which admit evolutionary fluctuations
in the Cartan system, permit the Topological Parity 4-form to be computed for
the Navier Stokes fluid; the result is:%

\begin{equation}
K=F\symbol{94}F=-2\nu\,(curl\,\mathbf{V}\circ curlcurl\,\mathbf{V)}%
dx\symbol{94}dy\symbol{94}dz\symbol{94}dt.
\end{equation}

From this result it is apparent that the Pfaff dimension of the domain is 4,
unless the viscosity is zero, or the vorticity field satisfies the conditions
of Frobenius integrability. The Torsion current anomaly is equal to
$-2\nu\,(curl\,\mathbf{V}\circ curlcurl\,\mathbf{V)}$. The torsion lines can
stop or stop within the domain producing defect structures that effect the
cohomology of the Cartan topology.

\qquad An interesting result is the proof that the closed integral of
topological Torsion-Helicity is a relative integral invariant for the viscous,
compressible fluid, if the Cartan sequence has a Pfaff dimension equal to 3.
Recall that the evolution of the 3-form $H=A\symbol{94}dA$ is given by the Lie
derivative expression,%

\begin{equation}
L_{(\beta V)}%
{\textstyle\oint_{3}}
H=%
{\textstyle\oint_{3}}
\{i(\beta V)dH+d(i(\beta V)H\}=%
{\textstyle\oint_{3}}
\{i(\beta V)dH\}+0
\end{equation}

But if $curl\,\mathbf{V}\circ curlcurl\,\mathbf{V}$ vanishes (for any
viscosity) then $dH=dA\symbol{94}dA=0$, and the RHS of the above expression
vanishes, for any reparameterization, $\beta$. Therefore, the closed integral
of the Topological Torsion three form is a deformation invariant not only of
Eulerian flows, but also of viscous flows for which the vorticity field is of
Pfaff dimension 2 (the velocity field is Pfaff dimension 3). The folklore
concept that viscosity destroys the helicity invariant is not necessarily true.

\medskip

\section{\bigskip Acknowledgments}

This article was presented as a talk given in August, 1991, at the Pedagogical
Workshop on Topological Fluid Mechanics held at the Institute for Theoretical
Physics, Santa Barbara UCSB.

\section{REFERENCES}

\lbrack1\rbrack\ E. Cartan ''Sur certaines expressions dfflerentielles et le
systeme de Pfaff'' Ann Ec. Norm. \textbf{16} 329 (1899)

\lbrack2\rbrack\ E. Cartan, ''Systems Differentials Exterieurs et leurs
Applications Geometriques'', Actualites sci. et industrielles 944 (1945)

[3] E. Cartan, ''Lecons sur la theorie des spineurs'' (Hermann, Paris 1938)

[4] E. Cartan, ''La Theorie des Spaces a Connexion Projective'', (Hermann,
Paris, 1937)

\lbrack5\rbrack\ S.S.Chern, Annais of Math. 45, 747- 752 (1944).

\lbrack6\rbrack\ R. M. Kiehn, ''Retrodictive Determinism'' Int. Journ. Eng Sci (1976)

\lbrack7\rbrack\ R.M. Kiehn, 'Topological Torsion, Pfaff Dimension and
Coherent Structures'' in Topological Fluid Mechanics, H. K. Moffatt and A.
Tsinober, editors, (Cambridge University Press, 1990), p. 225.

\lbrack8\rbrack\ H.Flanders, ''Differential Forms'', (Academic Press, N. Y. 1963).

\lbrack9\rbrack\ R.M. Kiehn and J. F. Pierce, The Physics of Fluids \textbf{9}
1941 (1969)

\lbrack10\rbrack\ R. M. Kiehn, J. Math Phys. 9 1975

\lbrack11\rbrack\ R. M. Kiehn, ''Are there three kinds of superconductivity''
INt J. of Mod. Phys.10 1779 (1991)

\lbrack12\rbrack\ W. Slebodzinsky, ''Exterior Forms and their Applications

\lbrack13\rbrack\ Van der Kulk and Schouten ''Pfaffs Problem Oxford University Press.

\lbrack14\rbrack\ S. Lipschutz, ''General Topology'', (Schaums Publishing
Co.,New York, 1965) p.97

\lbrack15\rbrack\ R. Hermann, ''Differential Geometry and the Calculus of
Variations'', (Academic Press, New York, 1968).

\lbrack16\rbrack\ R. M. Kiehn, Lett al Nuovo Cimento \textbf{14}, 308 (1975)
~Submersive Equivalence Classes for Metric Fields''

\lbrack17\rbrack\ W. Gellert, et.al. Editors 'The VNR Concise Encyclopedia of
Mathematics'', (Van Nostrand, New York 1977), p.686.

\lbrack18\rbrack\ J. G. Hocking, ''Topology'', (Addison Wesley, N. Y., 1961), p.2.

\lbrack19\rbrack\ R. L Bishop and S. l. Goldberg, ''Tensor Analysis on
Manifolds'', (Dover, N. Y., 1968).

\lbrack20\rbrack\ R. M. Kiehn, 'Topological Parity and the Turbulent State''
submitted to Jap. J. of Fluid Res.

\lbrack21\rbrack\ N. E. Kochin, l. A. Kibel, and N. V. Roze ~Theoretical
Hydrodynamics'' (Inter-science, New York 1965)

\lbrack22\rbrack\ H. W. Turnbull, ~''The Theory of determinants, matrices and
invariants'' (Dover, New York 1960)

\lbrack23\rbrack\ D. Struik, ''Differential Geometry''~, Addison Wesley,
(Reading, Mass 1961)

\lbrack24\rbrack\ J. Yorke and T.U, Amercan Mathematical Monthly \textbf{82},
985 (1975).

[25]\ R. M. Kiehn, Lett al Nuovo Cimento \textbf{12}, 300 (1975); Lett al
Nuovo Cimento \textbf{22}, 308 (1978)%

\begin{figure}
[ptb]
\begin{center}
\includegraphics[
height=6.6115in,
width=4.3405in
]{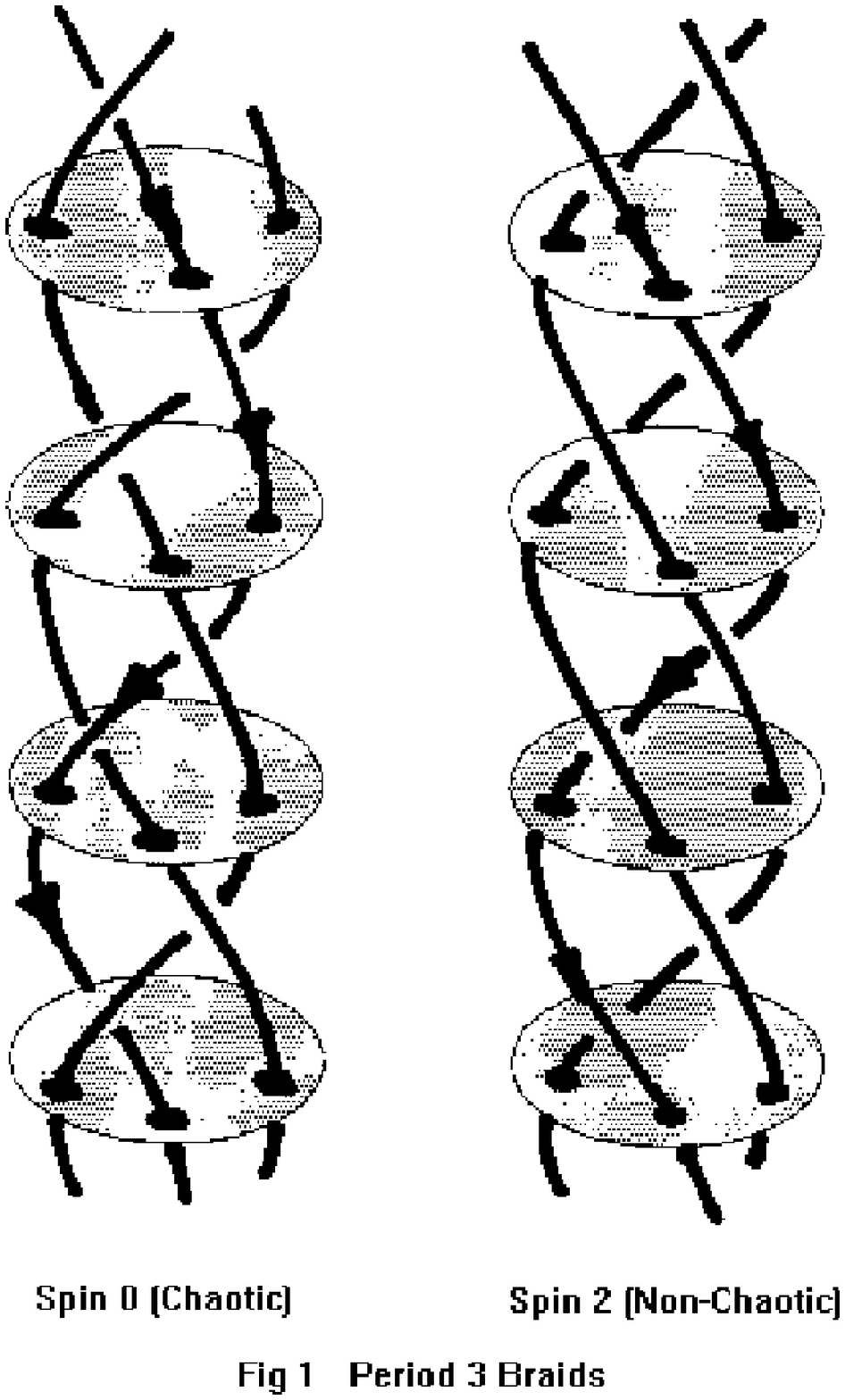}
\end{center}
\end{figure}
\end{document}